\documentclass[twocolumn,amsmath,amssymb,prl,superscriptaddress,fleqn]{revtex4-1}
\usepackage{epsfig}
\usepackage{graphicx}
\usepackage{dcolumn}
\usepackage{bm}
\usepackage{color} 
\newcommand{\fs}[1]{\textcolor{black}{#1}}
\usepackage{ulem}
\usepackage{amsmath}

\begin{document}


\title{Impact of Anisotropy on Antiferromagnet Rotation in Heusler-type Ferromagnet/Antiferromagnet Epitaxial Bilayers}

\author{T. Hajiri}
 \altaffiliation[Electronic mail:~]{t.hajiri@numse.nagoya-u.ac.jp}
\affiliation{Department of Crystalline Materials Science, Nagoya University, Nagoya 464-8603, Japan}
\author{M. Matsushita}
\affiliation{Department of Crystalline Materials Science, Nagoya University, Nagoya 464-8603, Japan}
\author{Y. Z. Ni}
\affiliation{Department of Crystalline Materials Science, Nagoya University, Nagoya 464-8603, Japan}
\author{H. Asano}
\affiliation{Department of Crystalline Materials Science, Nagoya University, Nagoya 464-8603, Japan}
\date{\today}
%
\begin{abstract}
We report the magnetotransport properties of ferromagnet (FM)/antiferromagnet (AFM) Fe$_2$CrSi/Ru$_2$MnGe epitaxial bilayers using current-in-plane configurations. 
Above the critical thickness of the Ru$_2$MnGe layer to induce exchange bias, symmetric and asymmetric curves were observed in response to the direction of FM magnetocrystalline anisotropy.
Because each magnetoresistance curve showed full and partial AFM rotation, the magnetoresistance curves imply the impact of the Fe$_2$CrSi magnetocrystalline anisotropy to govern the AFM rotation. 
The maximum magnitude of the angular-dependent resistance-change ratio of the bilayers is more than an order of magnitude larger than that of single-layer Fe$_2$CrSi films, resulting from the reorientation of AFM spins via the FM rotation. 
These results highlight the essential role of controlling the AFM rotation and reveal a facile approach to detect the AFM moment even in current-in-plane configurations in FM/AFM bilayers.
\end{abstract}


\maketitle

\section{INTRODUCTION}
Antiferromagnets (AFMs) show great potential to replace ferromagnets (FMs) in spintronic applications~\cite{AFM_spintronics, AFM_spintronics2, AFM_spintronics3}. 
Compared with FMs, AFMs have the advantages of much faster spin dynamics~\cite{Mn2Au_PRL, Nature_TmFeO3}, more stability against charge and external field perturbations~\cite{stability_AFM}, and no stray field~\cite{stray_1, stray_2}. 
However, since the AFM spins align in alternating directions of magnetic moments on individual atoms, the resulting zero net magnetization makes hard to control AFM magnetic moments.
Recently, there have been several reports regarding the control of AFM moments by applying an electronic current in AFM films~\cite{Sience_CuMnAs} and FM/AFM bilayers~\cite{Sakakibara_bilayer, CIMS_1}, by field cooling (FC)~\cite{FeRh_NatMat, AFM_FC} and by applying an external field via the exchange-spring effect~\cite{exchange_spiring, TAMR_Nat_Mat, SIO-LSMO, Hex_spring}. 
These studies demonstrated that the AFM moments can be controlled and detected using electronic transport measurements without the need for large-scale facilities such as synchrotron and neutron facilities~\cite{TAMR_PRL}.

AFM rotation is of interest because a more than 100\% spin-valve-like signal has been achieved in tunneling anisotropic magnetoresistance (TAMR) stacks by controlling the AFM spin configuration via the exchange-spring effect of FM on AFM~\cite{TAMR_Nat_Mat}. 
The exchange coupling has been widely used in spintronic devices such as spin-valve-type magnetic memory devices to pin the FM magnetization~\cite{Spintronics_1, Spintronics_2}. 
In contrast, TAMR utilizes the rotating AFM exchange-coupled to FM~\cite{rotatable_UC, exchange_spiring, XMCD_1}. 
The rotating AFM can be linked to the shift of hysteresis loops (exchange bias) and broadening of the coercivity in magnetization measurements~\cite{TAMR_PRL}. 
Although several studies have been reported regarding the rotating AFM~\cite{TAMR_Nat_Mat, TAMR_PRL, PMA_TAMR, YIG_IrMn}, almost all of the studies have been performed on polycrystalline stacks using AFM for IrMn. 
Since the AFM moments rotate with exchange-coupled FM, the AFM rotation behavior is expected to be affected by the FM magnetization switching process. 
Thus, the effect of FM magnetocrystalline anisotropy resulting from the full epitaxial growth is more interesting. 
In addition, all the studies have been performed using typical $3d$ metal FMs such as NiFe and Co. 
Similar to successful studies on giant magnetoresistance (GMR) and tunneling magnetoresistance (TMR)~\cite{GMR, TMR}, there is a clear need to study high-quality heterostructures with more advanced compounds such as those with high-spin polarization needed for spintronic devices~\cite{SPES_CMS}. 

For the advanced materials, we focused on the Heusler compound Fe$_2$CrSi (FCS), which is theoretically expected to be half-metallic FM~\cite{FCS_theory}. 
The four-fold magnetocrystalline anisotropy constant of FCS is known to be 266~J/m$^3$~\cite{Miyawaki_FCS}. 
In addition, since only fully epitaxial stacks effectively provide the advanced properties, the Heusler compound Ru$_2$MnGe (RMG) was selected for AFM; we succeeded in growing fully epitaxial FCS/RMG bilayers~\cite{Fukatani_bilayer}. 
RMG exhibits the highest N$\rm{\acute{e}}$el temperature, $T_N=353$~K, among Heusler compounds~\cite{Fukatani_RMG}. 
In addition, RMG has a nearly half-metallic electronic structure~\cite{RMG_HM}. 
Since the TAMR depends on spin configuration of AFM, such an AFM electronic structure is interesting.

In this letter, we systematically studied the magnetic and magnetotransport properties of FCS/RMG bilayers using current-in-plane (CIP) configurations. 
The angular-dependent resistance change ($\Delta R$) ratio and exchange bias ($H_{ex}$) exhibit a similar RMG thickness ($t_{\rm RMG}$) dependence, indicating that RMG spin reorientation via FCS rotation is dominant in $\Delta R$ above a critical thickness ($t_c$) to induce $H_{ex}$ even in CIP configurations. 
Above $t_c$, the magnetoresistance (MR) curves along the hard axes of FCS magnetocrystalline anisotropy exhibit a full rotation of AFM moments.
However, a partial rotation of AFM moments is observed along the easy axes, demonstrating the effect of the FM magnetization reversal process on AFM rotation. 
Although previous studies have used current-perpendicular-to-plane configurations, these results indicate that CIP magnetotransport measurements in FM/AFM bilayer provide a facile approach to detect AFM moments and promote their application in AFM spintronics.

\begin{figure}[b]
\includegraphics[width=7cm,clip]{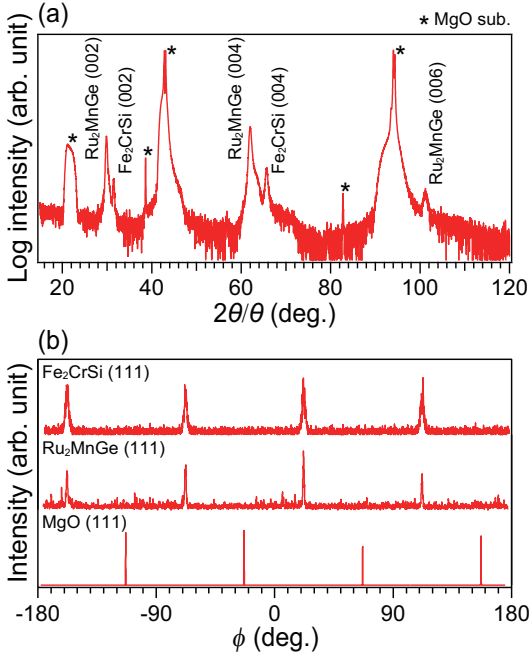}
\caption{
(Color online) Out-of-plane 2$\theta$/$\theta$ scans (a) and in-plane $\phi$-scans of FCS/RMG bilayers.
}
\label{fig:one}
\end{figure}

\section{EXPERIMENTAL DETAILS}
RMG/FCS bilayers were deposited on MgO (001) substrates by DC magnetron sputtering at a base pressure of approximately $5 \times 10^{-8}$~Torr. 
RMG thin films were deposited at a substrate temperature of $T_s=500$~$^\circ$C and then cooled to room temperature. 
Next, FCS was deposited on RMG at room temperature. 
After the FCS was deposited, the FCS/RMG bilayers were annealed at 500~$^\circ$C for 30~minutes to achieve $L2_1$ ordering of the FCS.
In addition, we also deposited FCS at $T_s=500$~$^\circ$C. 
The results were the same in both cases. 
The crystal structure was analyzed using both in-plane and out-of-plane X-ray diffraction (XRD) measurements with  Cu~$K\alpha$ radiation. 
The magnetic properties were characterized using vibrating sample magnetometry and superconducting quantum interference device (SQUID) magnetometry. 
The magnetotransport measurements were performed using the standard DC four-terminal method in the CIP configuration. 
To induce exchange coupling, the bilayers were annealed at 350~K for the SQUID measurements and at 375~K for the magnetotransport measurements for 30~minutes with applying field of +10~kOe, and then cooled to $T=4$~K with applying field of +10~kOe~\cite{Magnetic_treatment}.

\section{III. RESULTS AND DISCUSSIONS}

XRD patterns of the FCS/RMG bilayers are presented in Fig.~\ref{fig:one}(a) and~\ref{fig:one}(b). 
As observed in Fig.~\ref{fig:one}(a), only the $(00l)$ FCS and $(00l)$ RMG peak series show Bragg peaks in the out-of-plane XRD pattern. 
Epitaxial growth is confirmed by the in-plane $\phi$-scans presented in Fig.~\ref{fig:one}(b). 
Both RMG (111) and FCS (111) peaks are observed with shifts of 45$^\circ$ relative to the MgO (111) peaks.
These results indicate that their epitaxial relationship is FCS(001)[100]//RMG(001)[100]//MgO(001)[110]. 
In addition, since (111) reflection peaks of the Heusler alloy originate from superlattice reflections in the $L2_1$ ordered structure~\cite{Sakuraba_XRD}, these XRD results indicate that high-quality FCS/RMG bilayers were obtained.

\begin{figure}[b]
\includegraphics[width=8.5cm,clip]{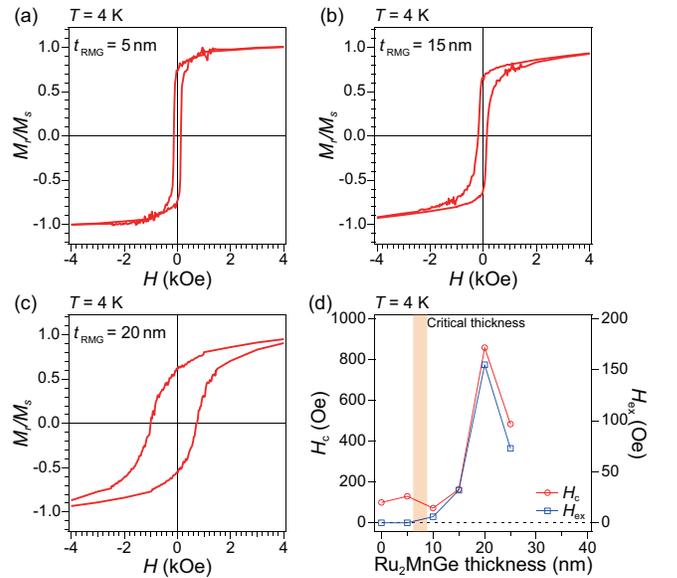}
\caption{
(Color online) (a)--(c) Magnetic hysteresis loops of $t_{\rm RMG}=5$,~15 and 20~nm at $T=4$~K after FC, respectively. 
(d) $H_c$ and $H_{ex}$ as a function of $t_{\rm RMG}$ at $T=4$~K.
}
\label{fig:two}
\end{figure}

\begin{figure*}[]
\includegraphics[width=17.5cm,clip]{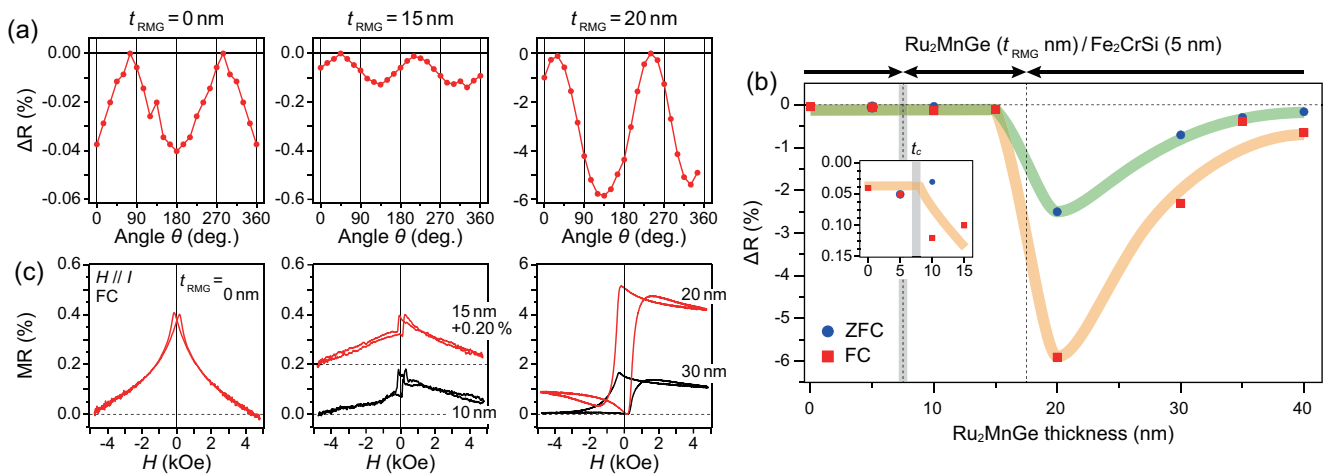}
\caption{
(Color online) (a) $\Delta R$ as a function of the angle between the current and field for single-layer FCS films and FCS (5~nm)/RMG ($t_{\rm RMG}$~nm) bilayers.
The measurements were performed at $T=4$~K with an applied field of +4~kOe.
(b) $\Delta R$ as a function of $t_{\rm RMG}$ after both ZFC and FC at $T=4$~K. 
(c) MR curves as a function of applied field on single-layer FCS films and FCS (5~nm)/RMG ($t_{\rm RMG}$~nm) bilayers at $T=4$~K.
}
\label{fig:three}
\end{figure*}

\fs{Figures~\ref{fig:two}(a)--\ref{fig:two}(c)} shows the magnetic hysteresis loops of FCS~(5~nm)/RMG~($t_{\rm RMG}$~nm) bilayers measured at $T=4$~K after FC. 
The measurements were performed along the easy axis of FCS $\langle100\rangle$. 
The $t_{\rm RMG}=5$ and 15~nm bilayers exhibit narrow hysteresis loops with a coercive field $H_c$ of approximately 100~Oe, which is similar to single-layer FCS films. 
The $t_{\rm RMG}=20$~nm bilayers exhibited a much wider hysteresis loop with $H_c$ of approximately 860~Oe. 
On the other hand, the $t_{\rm RMG}=5$~nm bilayer exhibited no hysteresis loop shift, whereas the $t_{\rm RMG}=15$~and~20~nm bilayers exhibited $H_{ex}=32$~and~155~Oe, respectively. 
The $t_{\rm RMG}$-dependent $H_c$ and $H_{ex}$ results are summarized in Fig.~\ref{fig:two}(d) at $T=4$~K. 
$H_{ex}$ is confirmed at above $t_{\rm RMG}=10$~nm, indicating that $t_c$ is between $t_{\rm RMG}=5$ and 10~nm. 
A maximum $H_{ex}$ appears at $t_{\rm RMG}=20$~nm; then, $H_{ex}$ decreases with increasing $t_{\rm RMG}$.
The same $t_{\rm RMG}$ thickness dependence was observed at $T=77$~K for a wider $t_{\rm RMG}$ range~\cite{supplement}.
$H_c$ shows no significant thickness dependence below $t_{\rm RMG}=15$~nm; then a jump is observed at $t_{\rm RMG}=20$~nm.
\fs{However, $t_{\rm RMG}=20$~nm shows much larger $H_c$ and $H_{ex}$ than $t_{\rm RMG}=25$~nm at $T=4$~K, of which results are unusual behavior.
The possible reason will be discussed later.}

Next, we focused on the $t_{\rm RMG}$-dependent magnetotransport properties \fs{using CIP configuration}.
Figure~\ref{fig:three}(a) plots the $\Delta R$ ratio as a function of the relative angle $\theta$ between the current and FM magnetization direction of FCS(5~nm)/RMG($t_{\rm RMG}~nm$) bilayers at $T=4$~K under an applied field of +4~kOe after FC.
At $t_{\rm RMG}=0$~nm, a typical anisotropic magnetoresistance (AMR) ratio is confirmed with a negative value of approximately -0.04 $\%$.
The negative AMR sign may originate from the half-metallic electronic structure of FCS, as discussed in recent theoretical and experimental studies~\cite{kokado_AMR, Sakuraba_AMR}. 
At $t_{\rm RMG}=15$~nm, the amplitude of $\Delta R$ increases, and its angular dependence shifts by approximately 45$^\circ$.
At $t_{\rm RMG}=20$~nm, the amplitude of $\Delta R$ is more than an order of magnitude larger than that of single-layer FCS films.
 
The $\Delta R$ ratios with respect to $t_{\rm RMG}$ at $T=4$~K after both zero-field cooling (ZFC) and FC are summarized in Fig.~\ref{fig:three}(b). 
After FC, the $\Delta R$ ratios are independent of $t_{\rm RMG}$ below $t_c$.
Above $t_c$, the $\Delta R$ ratios increase at 10 and 15~nm. 
At $t_{\rm RMG}=20$~nm, the $\Delta R$ ratio drastically increases and then decreases upon further increasing $t_{\rm RMG}$.
These $t_{\rm RMG}$-dependent $\Delta R$ ratios are similar to the exchange bias, as observed in Fig.~\ref{fig:two}(d). 
Moreover, as observed in Fig.~\ref{fig:three}(b), the $\Delta R$ ratios are enlarged by FC, demonstrating the effect of exchange coupling on the $\Delta R$ ratio.
Note that since the exchange coupling might exist even without FC~\cite{ZFC_Hex1, ZFC_Hex2}, the $\Delta R$ ratio after ZFC is larger than that for single-layer FCS films.

In addition, a relationship is observed between the $\Delta R$ ratios, the shape of the MR curves and exchange bias. 
The MR curves measured as a function of the applied magnetic field are presented in Fig.~\ref{fig:three}(c).
The MR curves of the single-layer FCS films ($t_{\rm RMG}=0$~nm~\cite{FCS_MR}) are symmetric, which originates from the FCS AMR. 
In contrast, above $t_c$, the curves of the bilayers differ from typical AMR curves.
The $t_{\rm RMG}=10$ and 15~nm bilayers exhibit small asymmetric curves, and the asymmetry increases at $t_{\rm RMG}=20$ and 30~nm.
According to previous FM/AFM studies~\cite{TAMR_Nat_Mat, TAMR_PRL, YIG_IrMn}, the asymmetric MR curves originate from the partial rotation of the AFM moments due to the applied external field via the FM rotation, whereas the symmetric MR curves originate from the full rotation of the AFM moments.

\begin{figure}[]
\includegraphics[width=8.5cm,clip]{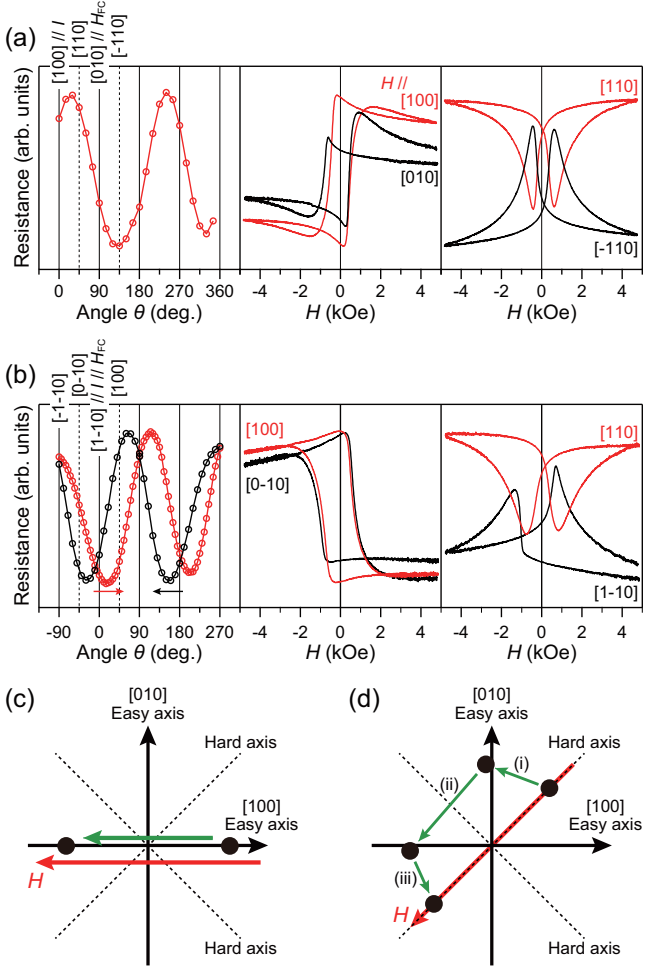}
\caption{
(Color online) (a, b) Magnetotransport properties of the configurations $I\parallel[100]$ and $\parallel[1-10]$, respectively. 
(c, d) FCS magnetization switching process where the field sweeps along the easy and hard axes, respectively.
}
\label{fig:four}
\end{figure}

Finally, we would like to discuss the origin of the anomalous magnetic properties of the bilayers.
Figures~\ref{fig:four}(a) and~\ref{fig:four}(b) compare different measurement conditions; the sensing current $I$ was applied in directions parallel to FCS/RMG [100] and [1-10], respectively.
As observed in Fig.~\ref{fig:four}(a) for $I\parallel[100]$, asymmetric MR curves are obtained along $H\parallel[100]$ ($\theta=0^\circ$) and [010] ($\theta=90^\circ$).
On the other hand, symmetric MR curves are obtained along $H\parallel[110]$ ($\theta=45^\circ$) and [-110] ($\theta=135^\circ$).
For $I\parallel[1-10]$, the symmetric and asymmetric relations with respect to crystalline direction do not change; symmetric MR curves are obtained along $H\parallel[1-10]$ ($\theta=0^\circ$) and [110] ($\theta=90^\circ$), and asymmetric MR curves are obtained along $H\parallel[100]$ ($\theta=45^\circ$) and [0-10] ($\theta=-45^\circ$).
Although FC-direction dependence was performed, no change was observed. 
These results indicate that the symmetric and asymmetric curves are not determined by the relative angle $\theta$ between $H$ and $I$, indicating that the obtained angular dependences are not typical AMR of FM.
Then, because the symmetric and asymmetric relations are not changed by the FC directions, the factor governing the angular dependence of the bilayers is not the sensing current or FC directions but the FCS/RMG crystalline direction.
 
One possible cause of the crystalline-direction-dependent AFM rotation is the FM magnetization switching process. 
FCS has four-fold magnetocrystalline anisotropy, where the easy axes are oriented along $\langle100\rangle$ and $\langle010\rangle$, and the hard axes are oriented along $\langle110\rangle$ and $\langle1-10\rangle$~\cite{Miyawaki_FCS}. 
This property indicates that symmetric MR curves appear along the hard axes of FCS, and asymmetric MR curves appear along the easy axes of FCS. 
It is well known that the FM magnetization switching processes with four-fold magnetocrystalline anisotropy differ along the easy and hard axes. 
As presented in Fig.~\ref{fig:four}(c), the magnetization rotates by 180$^\circ$ along the easy axes. 
On the other hand, as presented in Fig.~\ref{fig:four}(d), the magnetization rotates in 3 steps along the hard axes; (i) the magnetization rotates toward the nearest easy axis, (ii) the magnetization jumps in a direction close to the other easy axis, and (iii) the magnetization finally rotates toward the applied field direction~\cite{MOKE_PRB}.
Therefore, the magnetization rotates by up to 90$^\circ$ along the hard axis. 
A possible explanation for the full and partial AFM rotations along FCS/RMG $\langle100\rangle$ and $\langle110\rangle$ are that the AFM can fully follow the FM magnetization switching via the FM rotation when the field sweep is along the hard axis of FM because the magnetization rotates slightly (up to 90$^\circ$). 
On the other hand, the AFM cannot fully follow when the field sweep is along the easy axis of FM because the magnetization rotates a lot (180$^\circ$).
These results could provide a route to understanding the AFM rotation behavior.

Since the MR curves suggest the importance of AFM spin configuration, as discussed above, the larger $\Delta R$ ratio compared with that of only single-layer FCS films can be considered to be due to AFM moments.
To date, there have been several studies of AFM AMR due to spin flop~\cite{stability_AFM}, crystalline AMR originating from large anisotropies in the relativistic electronic structure~\cite{SIO-LSMO} and AFM spin configurations with respect to current direction~\cite{FeRh_NatMat}.
In the crystalline AMR study, the AFM spins were reoriented by applying magnetic fields via the exchange spring effect~\cite{exchange_spiring}, which is the same condition as that in our study. 
In addition,the $\Delta R$ ratio was obtained under $H=4$~kOe with small angle steps ($\sim$15$^\circ$), indicating that the AFM moments can fully follow the FM magnetization rotation as mentioned above. 
The AFM spin reorientation is reinforced by the delay of angular dependence due to the exchange-spring effect~\cite{TAMR_Nat_Mat}, as shown in Fig.~\ref{fig:four}(b). 
Then, the resistance is higher for the configuration of AFM moments aligned along [110] and [-1-10] than for the configuration of AFM moments aligned along [1-10] and [-110].
These resistance changes due to the AFM spin direction have been reported in FM/AFM bilayers~\cite{SIO-LSMO}. 
Moreover, the asymmetric MR curves were transformed into symmetric MR curves with increasing temperature.
Therefore, we conclude that the obtained larger $\Delta R$ ratio compared with that of only single-layer FCS films originates from the reorientation of the AFM moments by applying magnetic fields via FM rotation.
\fs{Then, the unusual $t_{\rm RMG}$-dependent $H_{ex}$ and $H_c$ might be caused by rotating AFM and/or exchange-spring effect.
As mentioned in the introduction, rotating AFM can be linked to $H_{ex}$ and $H_c$~\cite{TAMR_PRL}.
As similar to $H_{ex}$ and $H_c$, $t_{\rm RMG}=20$~nm shows much larger $\Delta R$ ratio than that of $t_{\rm RMG}=30$~nm.
Since $\Delta R$ ratio originates from the reorientation of the AFM moments via exchange-spring effect, these results might indicate the $t_{\rm RMG}$-dependent rotating AFM and/or exchange-spring effect.}
The obtained $\Delta R$ ratio of approximately 5.9~$\%$ is much larger than other AFM AMR ratios for Sr$_2$IrO$_4$ of approximately 1~\% and MnTe of approximately 1.6~\%.
This result might be related to either the nearly half-metallic RMG electronic structure or the electronic structures of both RMG and FCS.

\section{CONCLUSION}
We performed a magnetotransport study of FCS/RMG bilayers to clarify the AFM rotation behavior. 
In addition to the same $t_{\rm RMG}$ thickness dependence of the magnitude of the $\Delta R$ ratio and exchange bias, the MR curves changed from symmetric to asymmetric based on $t_{\rm RMG}$ in response to the direction of FCS magnetocrystalline anisotroy. 
The maximum $\Delta R$ ratio of the bilayers was more than an order of magnitude larger than that of single-layer FCS films due to the reorientation of AFM moments via the FM rotation. 
We also observed that the AFM moments could fully rotate when the field sweep was along the hard axes of FCS but could not fully rotate along the easy axes of FCS, demonstrating the impact of FCS magnetocrystalline anisotropy to govern the AFM rotation.
These results provide profound insights into the control of the AFM moments and promote the application of AFM in spintronics. 

\section{ACKNOWLEDGMENTS}
The authors gretefully acknowledge M. Kuwahara and K. Saitoh for their invaluable support.
Part of this work was supported by the Japan Society for the Promotion of Science (JSPS) Program for Advancing Strategic International Networks to Accelerate the Circulation of Talented Researchers.


\begin{thebibliography}{99}
\bibitem{Magnetic_treatment}\fs{The magnetic heat treatments have been done under the maximum temperature of each equipment of the SQUID measurements and the magnetotransport measurements.}
\end{thebibliography}
\end{document}